\documentclass{nature}

\usepackage{comment}

\usepackage{amsmath}
\usepackage[dvipsnames]{xcolor}
\usepackage{soul}
\usepackage{graphicx}
\usepackage{amssymb}
\usepackage{pifont}
\usepackage{longtable}
\usepackage{multirow}
\usepackage{arydshln}
\usepackage{hyperref}
\usepackage{caption}
\usepackage{pdflscape}

\usepackage[switch, modulo]{lineno}
\modulolinenumbers[1]

\newcommand{\kepler}{{\it Kepler}}

\newcommand{\RMS}{\mathrm{RMS}}

\newcommand{\cofiam}{{\tt cofiam}}

\newcommand{\multi}{{\sc MultiNest}}
\newcommand{\ultra}{{\sc UltraNest}}
\newcommand{\poly}{{\sc PolyChord}}
\newcommand{\emcee}{{\tt emcee}}
\newcommand{\luna}{{\tt LUNA}}
\newcommand{\pandora}{{\tt pandora}}
\newcommand{\wotan}{{\tt wotan}}
\newcommand{\satcand}{{\tt SatCand}}
\newcommand{\mlfriends}{{\tt MLFriends}}

\newcommand{\kep}{Kepler-1625}
\newcommand{\kepb}{Kepler-1625\,b}
\newcommand{\kepbi}{Kepler-1625\,b-i}
\newcommand{\gep}{Kepler-1708}
\newcommand{\gepb}{Kepler-1708\,b}
\newcommand{\gepbi}{Kepler-1708\,b-i}

\newcommand{\wwwcoolworlds}{\href{https://doi.org/10.5061/dryad.zs7h44jh4}{this URL}}

\newcommand{\youtube}{\href{https://youtu.be/v6Lb6OeSkZA}{this URL}}

\title{A Reply to: Large Exomoons unlikely around Kepler-1625\,b and Kepler-1708\,b}

\author{David~Kipping$^{1}\star$,
Alex~Teachey$^{2}$,
Daniel~A.~Yahalomi$^{1}$,
Ben~Cassese$^{1}$,
Billy~Quarles$^{3}$,
Steve~Bryson$^{4}$,
Brad~Hansen$^{5}$,
Judit~Szul\'agyi$^{6}$,
Chris~Burke$^{7}$,
Kevin~Hardegree-Ullman$^{8}$
}

\begin{document}

\maketitle

\begin{affiliations}
 \item Department of Astronomy, Columbia University, 550 W 120th St., New York, NY 10027, USA
 \item Institute of Astronomy and Astrophysics, Academia Sinica, Taipei 10617, Taiwan
 \item Department of Applied Mathematics and Physics, Valdosta State University, Valdosta GA, 31698, USA
 \item NASA Ames Research Center, Moffett Field, CA 94035, USA
 \item Mani Bhaumik Institute for Theoretical Physics, Department of Physics and Astronomy, UCLA, Los Angeles, CA 90095, USA
 \item Institute for Particle Physics \& Astrophysics, ETH Zurich, Wolfgang-Pauli-Str. 27, 8093 Z\"urich, Switzerland
 \item Department of Physics and Kavli Institute for Astrophysics and Space Research, Massachusetts Institute of Technology, Cambridge, MA 02139, USA
  \item Caltech/IPAC-NASA Exoplanet Science Institute, 770 S. Wilson Ave, Pasadena, CA 91106, USA
\end{affiliations}

\newpage

Recently, Heller \& Hippke\cite{HH23} argued that the exomoon candidates \kepbi\ and \gepbi\ were allegedly ``refuted''. In this Matters Arising, we address these claims. For \kepbi, we show that their Hubble light curve is identical to that previously published by the same lead author\cite{H19}, in which the moon-like dip was recovered. Indeed, our fits of their data again recover the moon-like dip with improved residuals than that obtained by Heller \& Hippke\cite{HH23}. Their fits therefore appear to have somehow missed this deeper likelihood maximum, as well producing apparently unconverged posteriors. Consequently, their best-fitting moon is the same radius as the planet, \kepb; a radically different signal from that which was originally claimed\cite{TK18}. The authors then inject this solution into the \kepler\ data and remark, as a point of concern, how retrievals obtain much higher significances than originally reported. However, this issue stems from the injection of a fundamentally different signal. We demonstrate that their Hubble light curve exhibits ${\simeq}20$\% higher noise and discards 11\% of the useful data, which compromises its ability to recover the subtle signal of \kepbi.

For \gepbi, it was claimed\cite{HH23} that the exomoon model's Bayes factor is highly sensitive to detrending choices, yielding reduced evidence with a biweight filter versus the original claim. We use their own i) detrended light curve and ii) biweight filter code to investigate these claims. For both, we recover the original moon signal\cite{K22}, to even higher confidence than before. The discrepancy is explained by comparing to their quoted fit metrics, where we again demonstrate that the Heller \& Hippke\cite{HH23} regression definitively missed the deeper likelihood maximum corresponding to \gepbi. We conclude that both candidates remain viable but certainly demand further observations.

\section*{Kepler\,1625b-i Specific Response}
We begin by first clearly stating: both exomoon candidates may not be real. Our original and continued claim is modest: these objects are \textit{candidates}, for which the data exhibits substantial but not entirely conclusive evidence in favour of exomoons.

Concerning \kep, a challenge in addressing their claims is that no machine-readable version of their reduced light curve was provided, unlike the original article\cite{TK18}, nor did they share these data following e-mail request. The authors also provide no description of their reduction of the Hubble data; a troubling omission given the notoriously large number of choices required to interpret an instrument with such strong systematics\cite{wakeford}. Nevertheless, we hypothesised that their reduction was the same as that published by the lead author previously\cite{H19}. As demonstrated in the SI and Figure~\ref{fig:K1625}, we find that the two light curves are indeed identical.

We compared the intra-orbit RMS (root-mean-square) of their Hubble light curve versus our own and another independent study of Kreidberg et al.\cite{K19} Excluding orbits 1 (settling), 7 (ingress) and 18 (egress), our original light curve\cite{TK18} has a median intra-orbit root mean square (RMS) of 361\,ppm. This becomes 362\,ppm for Kreidberg et al.\cite{K19} and 429\,ppm for Heller \& Hippke\cite{HH23} - indicating red noise at ${>}64\%$ the photon floor. They also drop the first exposure of every orbit\cite{H19}, leading to 11\% less data. The inferior precision in the Heller \& Hippke\cite{HH23} light curve immediately challenges its utility in reconstructing such a subtle signal as \kepbi.

Heller \& Hippke\cite{HH23} were unable to recover the moon-like dip tyat we reported\cite{TK18}. To investigate, we used \multi\cite{multi} to fit their light curve with a planet+moon model and the same long-term trend model as previously published\cite{H19}. We successfully recover the moon-like dip, but to lower significance than originally found due to their noisier reduction (see Figure~\ref{fig:K1625}). Our fit residuals are substantially lower than that presented in HH23 ($\RMS=422$\,ppm versus 459\,ppm), indicating that their fits missed this improved solution. In further support of this, we note that the lead author previously published an analysis with a different inference framework which successfully recovered the moon-like dip using the exact same data set\cite{H19}, yielding improved residuals of $\RMS=426$\,ppm. Together, this indicates that their new fits failed to recover the global maximum.

Heller \& Hippke\cite{HH23} argue that the evidence for \kepbi\ ``comes almost entirely from the shallowness of one transit observed with Hubble''. They ``interpret this as a fitting artefact in which a moon transit is used to compensate for the unconstrained stellar limb darkening''. Defining transit depth as the flux minimum, we agree that the Hubble light curve is shallower, which is an expected outcome of limb darkening when comparing visible and near-infrared bandpasses\cite{claret}. However, the \textit{geometric} transit depth, which corrects for limb darkening, is consistent using either data set (see SI). Crucially, both studies use the same limb darkening treatment and so this cannot explain our differing results. Further, comparison of our fitted limb darkening parameters with stellar theoretical models reveals good agreement (see SI). Thus, the depth difference is physically expected and does not induce such strange behaviour in our own fits.

We emphasise that Heller \& Hippke's\cite{HH23} planet+moon solution is radically different to that originally reported\cite{TK18}, which has major ramifications. The authors state that Figure~S1 lists their maximum likelihood parameters, where both $(R_{\mathrm{moon}}/R_{\star})$ and $(R_{\mathrm{planet}}/R_{\star})$ take a value of ${\sim}0.06$, whereas we reported\cite{TK18} $(R_{\mathrm{moon}}/R_{\star}) {\sim} 0.02$. They state that they injected this solution, essentially a binary planet, back into the \kepler\ data to infer the false-positive probability. This is less relevant, since the original claim rests upon the Hubble data, not \kepler. Regardless, in doing so, they find much higher Bayes factors ($2\log B_{mp} {\sim}100$-$1800$) in blind recoveries than we reported\cite{TK18} ($2\log B_{mp} {\sim}11$-$26$). Heller \& Hippke\cite{HH23} consider this a knock against the original claim, arguing that ``genuine [...] exomoons would tend to exhibit much higher Bayesian evidence'', whereas in fact their injection is fundamentally a different signal.

\section*{Kepler\,1708b-i Specific Response}
Heller \& Hippke\cite{HH23} report ``much lower statistical evidence for the exomoon candidate around \gepb'' than our original paper\cite{K22}. As with \kep, no machine-readable version of their light curve is provided. Their three detrending methods are described though: \#1 and \#2 using their own implementation of our \cofiam\ algorithm and \#3 using a biweight filter - all implemented within the \wotan\ package\cite{wotan}. Comparing to \#1 and \#2 is not instructive, since inspection of the \wotan\ code reveals its \cofiam\ is not equivalent (see SI) to our own (publicly available in ref. \cite{K22}). Regardless, as evident from their Figures 1 \& 3, the authors prefer the biweight filter and we use the same for consistency.

We ran \wotan\ with the biweight filter following the same options described by Heller \& Hippke\cite{HH23}, as well extracting the light curve presented in their Figure~3 (see SI). These light curves were fit with \multi\cite{multi} coupled to \luna\cite{luna} and clearly recover the same exomoon signal originally reported, to even higher significance (see Figure~2). Heller \& Hippke\cite{HH23} report that when using a planet-only model, the residuals exhibit an RMS of $529.9$\,ppm, which we also reproduce. However, for the planet-moon model, they report a residual RMS of $528.2$\,ppm, whereas we obtain a greatly improved $507.9$\,ppm. This conclusively demonstrates that their inference framework has again overlooked a deeper maximum, which may be due to insufficient steps used (see SI).

Heller \& Hippke\cite{HH23} report lower significances using \ultra\cite{ultra} and suggest the evidences we computed using \multi\cite{multi} are unreliable. To investigate this, we computed the Savage-Dickey ratio\cite{dickey} as an alternative (but approximate) means of computing the Bayes factor, yielding $\log(B_{mp})\simeq3.1$ (see SI), which is broadly consistent with the \multi\ result ($\log(B_{mp})=2.5$). Finally, we note that corner plots shown in the SI of Heller \& Hippke\cite{HH23} appear unconverged, unlike those presented originally (see Figure~\ref{fig:corners}). Taken together, the supposition that \multi\ is unreliable for this problem is not credible, and in fact we have demonstrated that the inference framework of Heller \& Hippke\cite{HH23} missed a deeper minimum - namely \gepbi.

Finally, they report a false-positive rate of 1.6\%, fully consistent our previously reported value of $1.0_{-1.0}^{+0.7}$\%. Heller \& Hippke\cite{HH23} suggest that red noise may have caused the exomoon signal, but this is fully encoded within this value and is self-evidently improbable. Further, applying the Ljung-Box, Box-Pierce and Kolmogorov–Smirnov tests reveals our original light curves are fully consistent with white noise (see SI).

\section*{Discussion}

Heller \& Hippke\cite{HH23} concluded that the exomoon candidates \kepbi\ and \gepbi\ are unlikely, but we have shown that their arguments are fundamentally flawed, stemming from numerous choices and interpretations that do not hold up to scrutiny. Further detailed point-by-point responses are provided in the SI. Nevertheless, skepticism about the reality of these candidates should be maintained, as we ourselves have previously voiced\cite{TK18,K22}. Accordingly, the exomoon hypotheses ultimately requires further observations to test.

\clearpage

\bibliographystyle{naturemag}
\newcounter{firstbib}


\begin{addendum} 
\item[Author Correspondence]
All correspondence regarding this work should be directed to D. Kipping.
\item
D.K. thanks donors Mark Sloan,
Douglas Daughaday,
Andrew Jones,
Elena West,
Tristan Zajonc,
Chuck Wolfred,
Lasse Skov,
Graeme Benson,
Alex de Vaal,
Mark Elliott,
Stephen Lee,
Zachary Danielson,
Chad Souter,
Marcus Gillette,
Tina Jeffcoat,
Jason Rockett,
Tom Donkin,
Andrew Schoen,
Jacob Black,
Reza Ramezankhani,
Steven Marks,
Nicholas Gebben,
Mike Hedlund,
Dhruv Bansal,
Jonathan Sturm,
Rand Corp.,
Leigh Deacon,
Ryan Provost,
Brynjolfur Sigurjonsson,
Benjamin Walford,
Nicholas De Haan,
Joseph Gillmer,
Emerson Garland,
Alexander Leishman,
The Queen Rd. Fnd. Inc.,
Brandon Thomas Pearson,
Scott Thayer,
Benjamin Seeley,
Frank Blood,
Ieuan Williams,
Jason Smallbon \&
Xinyu Yao.
D.K. \& D.A.Y. acknowledge support from NASA Grant \#80NSSC21K0960.
D.A.Y. thanks the LSSTC Data Science Fellowship Program, which is funded by LSSTC, NSF Cybertraining Grant \#1829740, the Brinson Foundation, and the Moore Foundation.
J.Sz. acknowledges the financial support from the European Research Council (ERC) under the European Union’s Horizon 2020 research and innovation programme (Grant agreement \#948467).
Analysis was carried out in part on the NASA Supercomputer PLEIADES (Grant \#HEC-SMD-17-1386), provided by the NASA High-End Computing (HEC) Program through the NASA Advanced Supercomputing (NAS) Division at Ames Research Center.
This paper includes data collected by the \kepler\ Mission. Funding for the \kepler\ Mission is provided by the NASA Science Mission directorate.
This work is based in part on observations made with the NASA/ESA Hubble Space Telescope, obtained at the Space Telescope Science Institute, which is operated by the Association of Universities for Research in Astronomy, Inc., under NASA contract NAS 5-26555. These observations are associated with program \#GO-15149. Support for program \#GO-15149 was provided by NASA through a grant from the Space Telescope Science Institute, which is operated by the Association of Universities for Research in Astronomy, Inc., under NASA contract NAS 5-26555.

\item[Author contributions] 
D.K. performed the data reduction, analysis and interpretation and wrote the majority of the text. A.T. performed much of the original TK18 analysis and consulted on all relevant sections here. B.C. benchmarked the \luna\ and \pandora\ codes. D.A.Y. used the \wotan\ code to compare light curve products. B.Q. performed the tidal migration analysis and wrote the corresponding SI section. B.H., S.B., J.S., C.B. \& K.H. helped edit the final version of the manuscript.
U.
\item[Author Information] Reprints and permissions information is available at www.nature.com/reprints. Correspondence and requests for materials should be addressed to D.K.~(email: dkipping@astro.columbia.edu).

\item[Competing Interests] The authors declare that they have no competing interests.

\end{addendum}


\newpage
\begin{figure}
\centering
\includegraphics[angle=0, width=16.0cm]{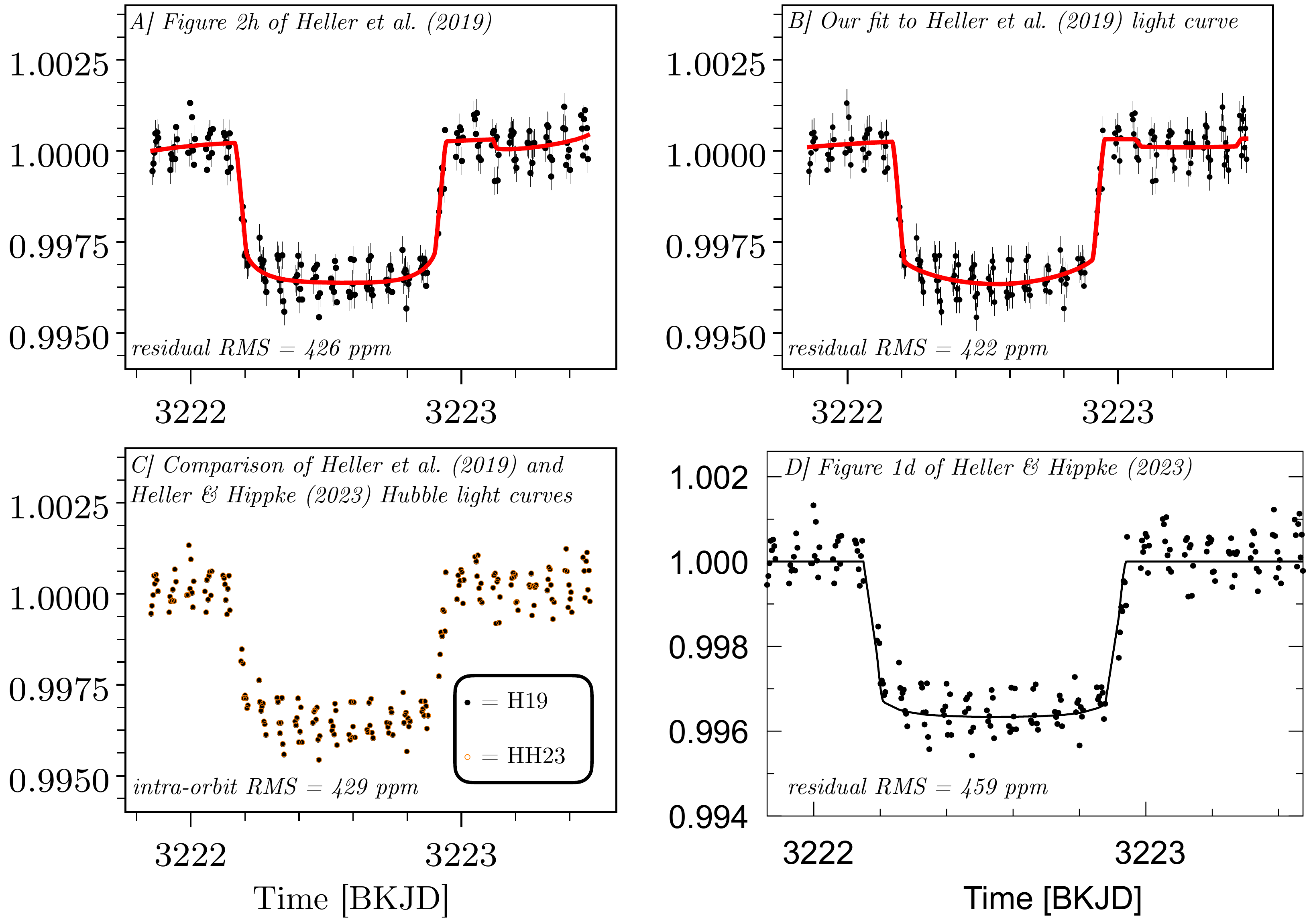}
\caption{\label{fig:K1625}
\textbf{Hubble light curves of \kepb.}
\textit{A:} Figure 2h from the 2019 publication of Heller et al.\cite{H19}, where we have removed the posterior samples to aid clarity. The moon-like dip is clearly recovered.
\textit{B:} Our own fits to the same data as A, where we have attempted to match the figure style for ease of comparison. The moon-like dip is again recovered, but with slightly improved residuals.
\textit{C:} Comparison of data in A versus the newly published Heller \& Hippke\cite{HH23} light curve, establishing they are identical.
\textit{D:} Figure 1d from Heller \& Hippke\cite{HH23}, where we have removed all lines except their best-fitting planet+moon solution. The residuals are now much higher than panel A or B, despite using the exact same light curve.
}
\end{figure}

\newpage
\begin{figure}
\centering
\includegraphics[angle=0, width=16.8cm]{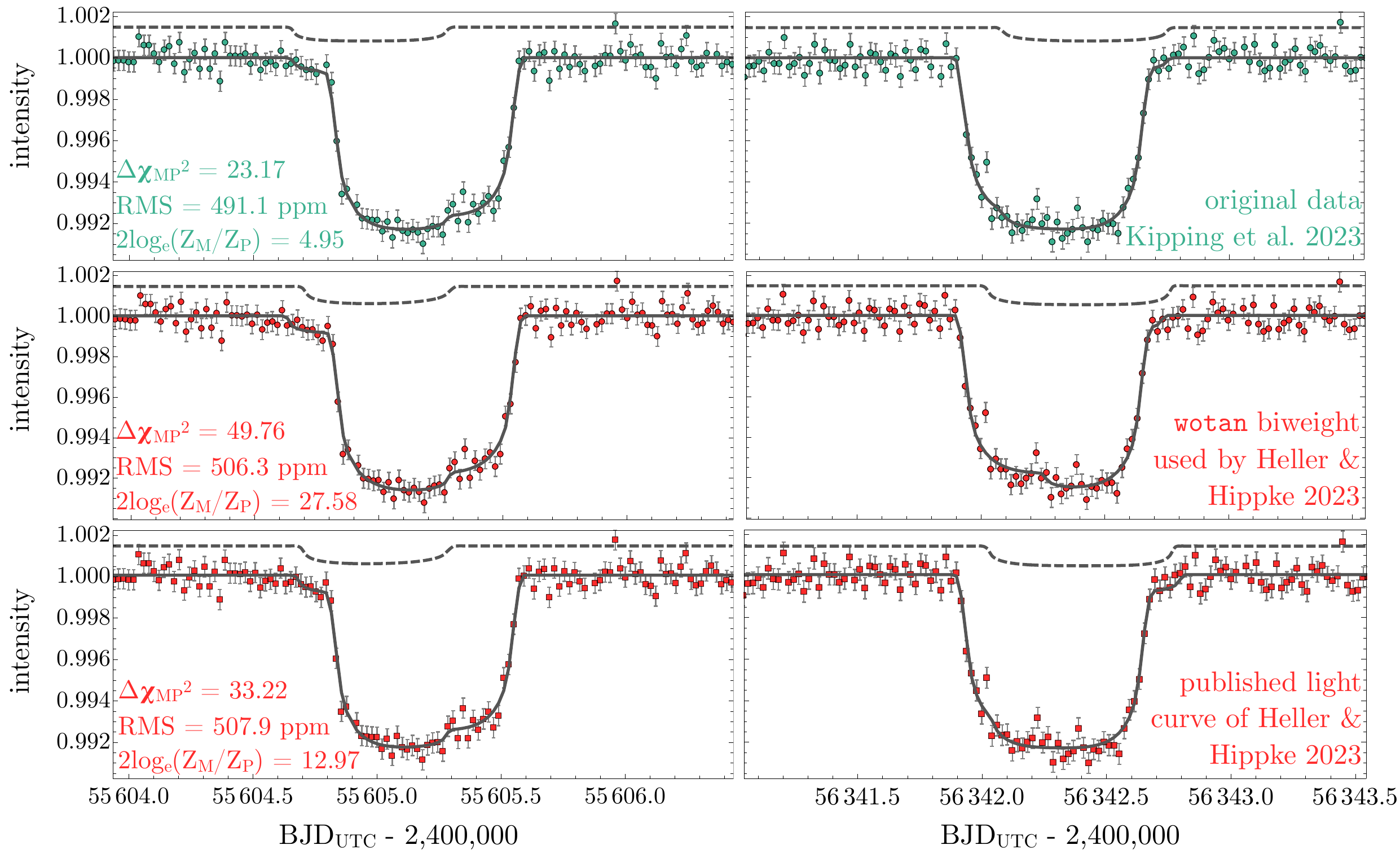}
\caption{\label{fig:K1708}
\textbf{\kepler\ light curves of \gepb.}
Top panel shows the two epochs (left \& right) of \gepb\ originally published\cite{K22}. The solid lines are the best-fit planet+moon model, and the dashed line is the isolated moon component. Row 2 shows the \kepler\ data processed with the \wotan\ biweight filter using the same options described in Heller \& Hippke\cite{HH23}, where we easily recover the same moon signal. Row 3 shows their extracted light curve, where we again easily recover the moon signal and obtain greatly improved residual RMS than that reported by Heller \& Hippke\cite{HH23} using their own data ($528.2$\,ppm).
}
\end{figure}


\clearpage

\noindent\textbf{\huge{Supplementary Information}}

\noindent\textbf{\large{Further Details Concerning Kepler 1625b}}
\label{si:k1625}

\noindent\textit{\textbf{K1625: Light curve extraction}}

Prior to the analysis of Heller \& Hippke\cite{HH23} (HH23 hereafter), three independent reductions of the \kepb\ Hubble observations had been published\cite{TK18,K19,H19}. No details are provided about the reduction of the Hubble light curve presented in HH23. However, inspection of the stochastic noise reveals familiar patterns to that presented by the same lead author in 2019\cite{H19} (H19 hereafter). No machine readable format of the light curve is provided in HH23, nor H19. The authors also did not share their light curve following our e-mail request. As a last resort, we downloaded the vector graphics source files uploaded to arXiv, which allow us to precisely reconstruct the flux values at each time stamp\footnote{Example available at \youtube}. This confirmed that the HH23 Hubble light curve is indeed identical to that of H19, as evident in Figure~\ref{fig:K1625}.

\noindent\textit{\textbf{K1625: The effect of bandpass on transit depths}}

HH23 highlight that the Hubble light curve of \kepb\ is significantly shallower than that observed by \kepler. One must be careful as to how one defines transit depth though, since multiple definitions exist. For example, it could be that calculated using the minimum flux, the mean in-transit flux or after correcting for limb darkening (geometric depth). HH23 focus their discussion on the first of these. The flux minimum is highly sensitive to stellar limb darkening, but this is fully accounted for standard light curve modelling\cite{mandel}. The fact that the Hubble flux minimum is shallower than \kepler\ is qualitatively what one would expect in going from a visible-bandpass instrument (\kepler) to a near-infrared one (Hubble WFC3). Thus, the qualitative observation of a shallower depth (using the flux minimum definition) was expected.

\noindent\textit{\textbf{K1625: Consistency of geometric transit depths between \kepler\ and Hubble}}

We argue that the more relevant metric to compare is the \textit{geometric} transit depth between the two bandpasses. We first establish that since the planet is massive and cool, we do not expect to see atmospheric absorption features. From reference \cite{winn}, the depth change expected due to atmospheric absorption is ${\sim}10 (R_P/R_{\star})^2 (H/R_P)$, where $H$ is the scale height. Assuming a 350\,K atmosphere under Jovian gravity, water vapour would induce a 3\,ppm feature, far below the ${\sim}70$\,ppm depth error found with Hubble\cite{TK18}. Accordingly, the geometric depth, defined as $(R_P/R_{\star})^2$, should be the same in both bandpasses.

Consider that this expectation was found to be false though - consider the scenario where the geometric transit depth of the Hubble light curve was shallower than that of \kepler. We label this ``scenario A''. This indeed would be a major problem. With atmospheric absorption out-of-play, the planet-only model has no way to explain such a difference (recall that limb darkening does not influence $(R_P/R_{\star})^2$). However, when attempting to fit a planet+moon model, the addition of the moon provides the extra freedom needed to explain away that depth difference. Notably, the fit could place the moon nearly on-top of the \kepler\ transits, thereby artificially inflating their depths. Indeed, HH23 reported that this situation describes approximately half of their posterior samples.

But that entire argument was predicated upon an unverified assumption - that the geometric transit depths were different between the two instruments. Let us now test that assumption. In fact, TK18 already did so, since in that work an additional parameter was introduced for the ratio of the Hubble light curve derived ratio-of-radii versus that from \kepler. For all three visit-long trend models attempted, the result was consistent with no geometric depth change (see their Table~2): $(R_{P,\mathrm{HST}}/R_{P,\mathrm{Kep}}) = (0.998\pm0.013)$ (linear), $(1.009\pm0.019)$ (quadratic) and $(1.006\pm0.014)$ (exponential). This scenario is thus not in effect for the TK18 light curve.

We repeated the exact same fits but replacing the TK18 light curve with the HH23 version and find $(R_{P,\mathrm{HST}}/R_{P,\mathrm{Kep}}) = (0.972\pm0.014)$, $(0.977\pm0.018)$ and $(0.974\pm0.014)$, respectively. Thus, there is modest evidence ($2$\,$\sigma$) that the HH23 Hubble light curve is shallower than the \kepler\ bandpass. Given that atmospheric absorption cannot explain such a difference, it may be indicative of a systematic error in the HH23 reduction itself.

Perhaps, then, scenario A causing a spurious moon might be valid for HH23's light curve, even if not true for TK18. But as seen later, even this does not appear to be true. Re-fitting the HH23 light curve using \luna\cite{luna} coupled to \multi\cite{multi} still recovers the moon-like dip, albeit to reduced significance which we attribute to their noisier reduction. This suggests that the difference may be attributed to the inference frameworks, which we will return to shortly.

\noindent\textit{\textbf{K1625: The benefit of allowing for a depth change}}

As noted earlier, TK18 allow for a geometric depth change between the \kepler\ and Hubble bandpasses. This was done for two reasons. First, the result allows us to rigorously check whether there exists any evidence for a depth difference. As already demonstrated, none was expected and thus the existence of any such depth change would be a red flag that something was at issue with the Hubble reduction. Second, even if an unexplained depth change did exist, by allowing for such a change in the code, it protects us against scenario A. It becomes far simpler for the model to explain the depth difference by simply adjusting the Hubble depth, rather than the contrived situation of placing a moon nearly precisely on top of the three \kepler\ transits yet also absent in the Hubble transit.

In Methods, ``Model parameterization'' of HH23, the authors note that TK18 allowed for a depth change yet found none, and then use this fact to justify not allowing for such a parameter in their own fits. However, since the authors are concerned about a depth change driving a spurious moon solution, this assumption seems in self-contradiction. In the end, just because the Hubble reduction of TK18 yields the expected result of no major wavelength dependency does not mean that they should assume their reduction passes the same test.

\noindent\textit{\textbf{K1625: Comparing limb darkening treatments}}

TK18 and HH23 follow the same prescription for limb darkening. Both use a quadratic limb darkening law parameterised by two freely fitted coefficients. Both assume that these coefficients are distinct and independent for the Hubble and \kepler\ bandpasses. The fact that the limb darkening treatments are identical rules this out as a possible explanation for why HH23 arrive at a different light curve solution.

\noindent\textit{\textbf{K1625: Physical consistency of the fitted limb darkening parameters}}

Perhaps the final issue to discuss regarding limb darkening is whether the fitted coefficients are actually physically sound. If scenario A were in effect, one might argue that the reason TK18 find $(R_{P,\mathrm{HST}}/R_{P,\mathrm{Kep}}){\sim}1$ is because the freely fitted limb darkening parameters are driven to extreme choices to promote that. This is already a contrived argument since the prior on $(R_{P,\mathrm{HST}}/R_{P,\mathrm{Kep}})$ is uninformative; there's no reason for the code to try and enforce it to be unity at the expense of limb darkening. Regardless, it can be fully resolved by inspection of the \textit{a-posteriori} limb darkening parameters and comparing to astrophysical models. Unfortunately, no tabulated models of limb darkening coefficients exist for the Hubble WFC3 G141 bandpass (to our knowledge), and thus it was necessary to calculate our own.

This was achieved by taking the reported effective temperature and $\log\mathrm{g}$ of the star\cite{TK18} and following the prescription described in reference \cite{bakos}, which uses the Kurucz stellar atmosphere \href{http://kurucz.harvard.edu/grids.html}{model database}. The only difference is that repeated the process once for the \kepler\ bandpass and a second time for the WFC3 G141 bandpass, available at the \href{https://hst-docs.stsci.edu/wfc3ihb/chapter-8-slitless-spectroscopy-with-wfc3/8-3-slitless-spectroscopy-with-the-ir-g102-and-g141-grisms}{HST WFC3 Instrument Handbook}. For the \kepler\ bandpass, this yields $q_1 = 0.479$ and $q_2 = 0.335$ and for WFC3 this yields $q_1 = 0.203$ and $q_2 = 0.154$. Extended Data Figure~\ref{fig:LDCs} compares these theoretical limb darkening coefficients with the posterior distributions reported from TK18, using their quadratic visit-long trend with the planet+moon model. As the figure reveals, both bandpasses appear consistent with theoretical expectation and there is no evidence that our limb darkening treatment is at issue.

\noindent\textit{\textbf{K1625: A lower precision Hubble light curve}}

We first note that HH23 (=H19) drop the first exposure of every orbit (26 in total). This means there is 11\% less useful data (196 data points rather than 220) than that used in TK18 and indeed K19.

We also note that the photometric precision is substantially worse in the Hubble light curve of HH23 (=H19). A useful measure here is the intra-orbit RMS. Hubble conducted 26 orbits around the Earth to collect its observations of \kep. Because the telescope is physically behind the Earth half of the time, the data features large gaps every orbital period of the spacecraft (95\,minutes). As a result, we obtain clusters of points of roughly 45-minute duration, each featuring a set of typically nine 5-minute exposures. Ignoring the planetary ingress/egress, we do not expect any sharp temporal astrophysical deviations on this 45-minute timescale and thus these clusters of points should be close to one another. Accordingly, for each Hubble orbit, we measure the RMS within that orbit, and then take the average of these values as a singular metric of precision. The advantage of this approach is that it's independent of the transit or visit-long trend model adopted by either party.

Excluding orbits 1 (settling), 7 (ingress) and 18 (egress), we measure a median intra-orbit RMS of 360.6\,ppm for TK18. Repeating for the independent reduction of K19, we measure a slightly worse score of 362.4\,ppm. Note that these are the same values reported previously\cite{loose}. For HH23 (=H19), we measure a median intra-orbit RMS of 428.5\,ppm - a much higher value. In fact, this implies an additional red noise component of ${>}64$\% the photon-floor (since noise adds in quadrature) and thus appears heavily contaminated. As an extra check, we also fit all three data sets with the three visit-long trend models from TK18 (linear, quadratic, exponential) with the planet+moon model and measured the residual RMS. As before, this reveals a sizeable gap in precision between the HH23 (=H19) light curve and that of TK18 and K19 (see Extended Data Table~\ref{tab:K1625}).

We note that H19 used an exponential hook correction for the ramp effect within each orbit. TK18 also tried this but found that a novel non-parametric method (introduced in that work) was far superior in terms of final precision. We note that K19 follow the same non-parametric correction procedure as devised in TK18. We hypothesised that this choice could explain the differences seen and thus computed the median intra-orbit RMS of the TK18 exponential hook-corrected light curves, giving 396.9\,ppm. Thus, even repeating their (less precise) hook correction approach, we are unable to replicate such a high noise value as that obtained in HH23, which again challenges its utility in assessing the reality of \kepbi.

\noindent\textit{\textbf{K1625: Recovering \kepbi\ in the HH23 light curve}}

We took the HH23 Hubble light curve and combined it with the method marginalised \kepler\ light curve\cite{TK18} to re-fit the joint set using the same algorithms used in TK18. This is in fact represents the third fit conducted on the HH23/H19 light curve. H19 produced the first such fit, in which they used \emcee\cite{emcee} coupled to a planet+moon forward model published in reference \cite{rodenbeck} as well as a cubic polynomial for the visit-long trend. In that work, which we note uses the exact same data and indeed has the same lead author as HH23, the moon-like dip is recovered in the same location as that originally reported in TK18 (see Figure~1h of H19). H19 also reported remarkably similar properties between their fits of their own Hubble reduction versus the TK18 Hubble light curve, as evident from their Table~1. For example, the moon's semi-major axis is reported as $2.9_{-1.0}^{+1.5}$ stellar radii, versus $2.9_{-0.6}^{1.3}$ using TK18. Likewise, H19 report an exomoon radius of $0.016_{-0.005}^{+0.005}$ stellar radii, versus $0.019_{-0.005}^{+0.005}$ using TK18. Thus, by all accounts, it is the signal of the same exomoon candidate. Although H19 do not quote the RMS residuals of their best-fit, shown in their Figure~2h, our extraction reveals it to be 426.3\,ppm - which is close the median intra-orbit RMS (see Table~\ref{tab:K1625}), indicating a satisfactory fit.

The second fit of this data set is next reported in HH23. Here, the inference changes to \ultra\cite{ultra} coupled to \pandora\cite{pandora} for the planet+moon model. The authors do not report the RMS scores of their residuals for the Hubble data in isolation, but extracting the residuals from their Figure~1l we measure it to be 458.7\,ppm. With the same data set then, the revised fits of HH23 obtain \textit{worse} residuals than that of H19 (426.3\,ppm), and indeed higher than expected given the intra-orbit RMS of 428.5\,ppm. HH23 state that they are unable to recover the same moon-like dip. Given that the H19 Hubble light curve is identical to the HH23 light curve, this begs the question as to why.

As a third fit on the H19(=HH23) Hubble light curve, we performed our own regression using \multi\cite{multi} coupled to \luna\cite{luna}. There still remains the question as to what visit-long trend model to adopt, but to keep things as equal to H19 as possible, we follow their prescription of a cubic polynomial. In doing so, our fits once again recover the same moon-like dip claimed in TK18, and indeed also reported in H19 (see Figure~\ref{fig:K1625}). This results in a residual RMS of 422.0\,ppm, which is comparable to that of H19 (426.3\,ppm), but clearly much lower than that found by HH23 (458.7\,ppm). We note that the significance of the moon-like dip is diminished, with $2\log B_{MZ} = (1.64\pm0.32)$ and $\Delta\chi_{MZ} = 3.25$, when comparing to a zero-radius moon model - however, this is broadly to be expected given their noisier reduction and smaller data set. We again emphasise that all three fits are on the exact same Hubble light curve.

The only explanation is that the inference framework adopted in HH23 is unable to identify a deeper likelihood maximum within the parameter volume. This explains both why H19 previously found the moon-like dip using \emcee\cite{emcee}, and why our own fits of their light curve obtain lower RMS residuals.

Further doubt is cast on the HH23 inference framework by inspection of their posterior corner plot, which they provide in their SI, revealing what appears to be unconverged distributions. This behaviour is not seen using \multi\cite{multi}. A like-for-like comparison is provided in Extended Data Figure~\ref{fig:corners}, where we argue this is self-evident.

\noindent\textit{\textbf{K1625: Conclusions}}

We have established the following points that we believe explain the differing results of HH23 to TK18.

\begin{enumerate}
\item The HH23 and H19 Hubble light curves are identical and this light curve is noisier than all previously published reductions, comprised of ${>}64$\% red noise, as well as featuring 11\% less data.
\item The \kepler\ and Hubble light curves are consistent with no geometric depth change and physically sensible limb darkening effects, which is fully accounted for in the original analysis.
\item The HH23 inference framework is unable to recover a deeper likelihood maximum corresponding to the moon-like dip originally reported in TK18. This dip is recovered on their own Hubble light curve both in the lead author's previous work\cite{H19} and again here.
\end{enumerate}

\noindent\textit{\textbf{K1625: Point-by-point rebuttal}}

\noindent We here address, point-by-point, the critiques raised in HH23 in the case of \kepbi.

\noindent\textit{
``(1) About half of the posterior models do not exhibit a single moon transit in any of the four transit epochs...''}
This behaviour is indeed strange, but the likely explanation is that their inference framework has failed to converge to the global likelihood maximum, as demonstrated in our earlier analysis of their light curve. This behaviour is categorically not seen in the TK18 posteriors. For example, in the quadratic visit-long trend model fits, over 98\% of the posteriors include a moon transit in one of the four transits. We emphasise that our posteriors were made fully public in TK18 and thus could have easily been checked.

\noindent\textit{
``(2) In the other half of our posterior models that do contain moon transits, these transits occur almost exclusively in the Kepler data...''}
Same response as point (1).

\noindent\textit{
``(3) From these posterior cases with a moon transit, we find only a handful of light curves with a notable out-of-planetary-transit signal from the moon...''}
Same response as point (1).

\noindent\textit{
``(4) ...We did not find any evidence of a putative exomoon signal at 3,223.3 d (BKJD) in the Hubble data (Fig. 1d) as originally claimed...''}
This statement contradicts the lead author's own previous analysis of the exact same light curve where the moon-like dip was recovered\cite{H19}. Further, our own independent fits of their light curve recover the moon-like dip (see Figure~\ref{fig:K1625}). Both of these exhibit lower residuals than that presented in HH23.

\noindent\textit{
``(5) The transit observed with Hubble is much shallower than the three transits observed with Kepler.''}
Hubble WFC3 is a near-infrared instrument, whereas \kepler\ is operates in the visible bandpass. Limb darkening is suppressed towards the infrared and thus naturally explains why the flux minimum appears shallower\cite{claret}. The real question is whether the change is consistent within the limits of physical limb darkening. We have established that this indeed true by calculating the theoretically expected limb darkening coefficients and comparing to those found from the light curve fits. Further, the limb darkening treatment in HH23 and TK18 is the same, so this cannot explain the differing results.

\noindent\textit{
``(6) ...It is the transit depth discrepancy that causes the spurious moon signal.'''}
Same response as point (5).

\noindent\textit{
``(7) ..this metric for the noise amplitude [RMS] is larger than the depth of the claimed moon signal of about 500 ppm.''}
An RMS value in isolation cannot be used to assess the signal-to-noise (S/N) of a signal. The RMS is computed over the native cadence, here one exposure every 5\,minutes. The relevant noise metric is not the RMS, but rather the \textit{time integrated} noise\cite{SNR}. From TK18, the moon-like dip lasts from 3223.14\,BKJD until the end of the Hubble observations and thus spans 48 data points. Accordingly, the RMS should be divided by square root 48 to obtain the relevant noise scale, which using HH23's value of 619.1\,ppm would yield 89.1\,ppm. This is the value which should be compared to the 500\,ppm depth of the moon-like dip in discussions of detectability. We further note that HH23 appear to have taken the RMS of the residuals of the joint \kepler\ and Hubble light curves, which are highly heteroskedastic with respect to one another and thus this chimera RMS is not particularly instructive in any case.

\noindent\textit{
``(8) Our properly phase-folded exomoon transit light curve has a marginal S/N of only 3.4 or 3.0 depending on the detrending. There is also no visual evidence for an exomoon transit in this phase-folded light curve of Kepler-1625 b.''}
TK18 report that the moon-like dip has an S/N of (see their Table~1) of 5.8, 4.4 and 4.4 for the linear, quadratic and exponential visit-long trend models. The fact that HH23 report an S/N of 3.0 to 3.4 is actually consistent with these values when one accounts for the fact their Hubble light curve is significantly noisier. For example, taking the ratio of the intra-orbit RMS values and multiplying by the quoted S/N of 3.4, and the reduction of 11\% less data, gives $(428.5.0/360.6) \times 3.4 \times \sqrt{1.11} = 4.3$. Therefore, the HH23 diminished S/N appears consistent with the original detection after accounting for their less precise Hubble reduction. We emphasise that ``visual evidence'' is inherently subjective and objective evaluations of the reality of a signal should be based upon rigorous statistical metrics\cite{childs}. More importantly, a phase-folded analysis is inappropriate since the number of transits (4) does not satisfy the criteria of being $\gg1$ necessary for averaging effects to come to bear\cite{origami}.

\noindent\textbf{\large{Further Details Concerning Kepler 1708b}}
\label{si:k1708}

\noindent\textit{\textbf{K1708: Reproducing the HH23 light curves}}

HH23 do not provide a machine readable version of their \gepb\ \kepler\ light curve (and we remind the reader there is no Hubble data for this target). The authors also did not share their light curve following our e-mail request. In order to reproduce their work, we follow their description of their method \#3 by running \wotan\ with the Tukey biweight filter. As closely as possible, we match the various inputs to that described by HH23, such as using a window size of three times the planetary transit duration. The final light curve appears ostensibly remarkably similar to the method marginalised light curve presented in Kipping et al.\cite{K22} (K22 hereafter), as evident in Figure~\ref{fig:K1708}. Indeed, we note that K22 found that all eight detrending methods produce very similar light curves.

There are numerous options within \wotan\ we were forced to guess, such as the precise value of the transit duration used and the transit ephemeris used for masking. For this reason, we believe our \wotan\ light curve is a close match to that presented in HH23, but not identical. To obtain a truly identical light curve, we again turn to the vector graphics figure source files provided in HH23's arXiv submission (their Figure~3). HH23 do not state how many data points were used in their \kepler\ light curve, but our extraction identified $n=385$ points. HH23 \textit{do} quote the RMS and RSS values from fits upon these data though, and we note that $n=\mathrm{RSS}/\mathrm{RMS}^2$. For both the planet-only and planet-moon fits, the HH23 quoted values imply $n=386$ points. We thus concluded a single data point was missing from their Figure~3. However, the time and flux value of this point is unknown. Although this single point is unlikely to affect things significantly, for the sake of completion we decided to reverse engineer its probable value.

We first note that epoch 1 has 193 points but epoch 2 and 192. Under the assumption HH23 used a fix-width interval filter for extracting the light curves, this implies the missing point is located in epoch 2. However, by comparing the time stamps of epoch 2 to the raw \kepler\ light curve, we can confirm there are no missing cadences and thus the missing point must either be just before the first point, or just after the last point.

Next, we fit the $n=385$ point light curve with a planet-only model and obtain RMS\,$=529.1$\,ppm and RSS\,$=107.8$\,ppm$^2$, both of which very close to values HH23 report of $529.9$\,ppm and an RSS of $108.4$\,ppm$^2$, respectively. Despite missing a data point, the excellent agreement between these values already establishes that our extraction of their light curve was accurate. 

To predict the flux value of the missing point, for both of its possible locations, we note the extraordinarily high correlation between the K22 and HH23 light curves, of 0.999386. Therefore, the K22 flux values provide an excellent approximation for the missing flux point. Despite this, very slight differences do exist between the two as a result of their differing detrending strategies. To account for this, we applied a second-order spline interpolation through the deviances between them and then extrapolate that function one cadence out in each direction to apply the appropriate correction. The result is two possible light curves. To choose between them, we reproduce the planet-only fit used by HH23 and calculate the residuals against the best-fit. Assuming the missing point is located before the second transit, yields an RMS\,$=528.5$\,ppm and RSS\,$=107.8$\,ppm$^2$, whereas placing it after the second transit gives RMS\,$=529.3$\,ppm and RSS\,$=108.1$\,ppm$^2$. For comparison, HH23 report an RMS of $529.9$\,ppm and an RSS of $108.4$\,ppm$^2$. On this basis, we conclude that the missing point is most likely located after the second transit. In what follows, we adopt that version as a our final reconstruction of the HH23 light curve. This light curve, as well as the \wotan\ biweight light curve, are presented in Figure~\ref{fig:K1708}.

\noindent\textit{\textbf{K1708: Comparison to the K22 light curve}}

It's instructive to compare the precision of the various light curve products. As already noted, the only quoted values in the HH23 to guide such a comparison are the root mean square (RMS) and residual sum of squares (RSS) values quoted in reference to planet-only and planet+moon fits conducted. Specifically, for the planet-only fit, HH23 report an RMS of $529.9$\,ppm and an RSS of $108.4$\,ppm$^2$. Using their extracted light curve, our own light curve fits with the planet-only model obtaining an RMS of $529.3$\,ppm and an RSS if $108.1$\,ppm$^2$ - i.e. almost identical. The goal of that earlier fit was to reproduce the procedure of HH23, which from their Figure~3 we surmised used a fixed baseline of unity. In contrast, the fits of K22 allow the baseline to be optimised at each step, following the procedure of reference \cite{kundurthy}. Repeating the fits with this minor change leads to a very slight improvement of RMS\,$=527.8$\,ppm and RSS\,$=107.5$\,ppm$^2$.

We now compare these numbers to that found using the original K22 light curve, obtained via method marginalised detrending\cite{K22}. For the planet-only fit, we obtained an RMS of $495.9$\,ppm, which is substantially lower than that obtained by HH23. This establishes that the K22 detrending is certainly more precise. We also repeated this exercise with the \wotan\ biweight filter light curve, which is again noisier, and present the results in Table~\ref{fig:K1708}.

\noindent\textit{\textbf{K1708: Evidence that HH23's search method overlooks the exomoon signal}}

As discussed in the main text, it appears that HH23's inference framework of \ultra\cite{ultra} and \pandora\cite{pandora} somehow misses a deep global minimum present - the proposed exomoon signal. This determination was based upon the following. We first begin with the HH23 extracted light curve. It has already been established that our extraction of the HH23 light curve was accurate, since the RMS and RSS values from our planet-only fits are almost identical to that quoted by HH23. With this in mind, we next used \luna\cite{luna} with \multi\cite{multi} to regress the planet+moon model to their data.

The maximum \textit{a-posteriori} solution is shown in the bottom panel of Figure~\ref{fig:K1708}, which is manifestly the same solution originally reported in TK18. Further, both the Bayes factor and $\Delta\chi^2$ values indicate even higher statistical significance for the moon solution than originally reported - for example here we find $2\log B_{mp} = 6.90$ versus the original $2\log B_{mp} = 4.95$. This finding is in stark contrast to that reported by HH23, who report (with the same data set) a marginal Bayes factor of $2\log B_{mp} = 2.8$. Given that the data set is essentially identical, the origin of this discrepancy can only be due to differing inference frameworks used.

This raises the question - which inference framework is right? HH23 suggest that it is that of TK18 which is at fault. The authors highlight that \ultra\ was designed to provide robust evidence values\cite{ultra} and thus seem to imply that it's results should be taken at face value as always superior. However, we caution that when in run step sampling mode, as HH23 chose to do so, this is not necessarily true since \ultra\ behaves like \poly\cite{poly} and is thus the reliability of the results are sensitive to the number of steps selected. \ultra's default sampling method is the \mlfriends\ algorithm, which is designed to provide very secure Bayes factors but it comes at much larger computational cost. Presumably, that high cost is why HH23 did not use it. 

To investigate the claim that our \multi\ evidence values are unreliable, we calculated the evidence using an alternative technique - the Savage-Dickey ratio\cite{dickey}. We ran a moon model where we turn off our requirement for a physically sensible moon density, thus permitting zero-radius moons. We also allow for negative radius moons in this fit (flipped moon transits) to remove any boundary condition at zero that could impact the sampler. Following the Savage-Dickey theorem, we can calculate the Bayes factor of a zero-radius moon versus a finite-radius moon by taking the ratio of the posterior density at zero radius versus the density of the prior. This takes advantage of the fact that with just two transits, it is not possible for a transit timing variation signal to drive a finite moon mass. In general, the Savage-Dickey method is less precise since it relies on samples at the tail of the distribution, which are necessarily subject to small number statistics. Nevertheless, the Savage-Dickey ratio yields $\log(B_{mp})\simeq3.1$, whereas \multi\ originally returned $2.5$ in K22. Accordingly, the values appear broadly consistent, at least for the planet+moon problem.

A much more impactful test is to compare the RMS values of the planet+moon fit residuals. HH23 report that their best fit solution yields an RMS of $528.2$\,ppm, marginally improved over the planet-only value of $529.9$\,ppm. Running our own inference framework on their light curve, we obtain a much lower RMS value from our best-fit solution of $507.9$\,ppm (see Figure~\ref{fig:K1708} and Extended Data Table~\ref{tab:K1708}). This establishes, beyond doubt, that the HH23 inference framework definitively missed a deep likelihood maximum. Naturally, missing such a deep maximum will lead to attenuated Bayes factors.

Further, inspection of their posterior corner plot, which they provide in their SI, reveals what appears to be unconverged distributions, unlike those produced by \multi. This is similar to the situation found with \kepbi\ earlier (see Extended Data Figure~\ref{fig:corners}).

\noindent\textit{\textbf{K1708: Injection-recovery tests of \multi\ + \luna}}

HH23 report that their injection recovery experiments for an exomoon injected around \gepb\ into the \kepler\ data produce much higher Bayes factors than that reported in K22; HH23 find $10 \lesssim 2 \log B_{mp} \lesssim 100$. The authors comment that this makes the real discovery very surprising amongst the landscape of possible Bayes factors. We would add a correction here that the extrema of the range are less important than the \textit{distribution} of those Bayes factors; extrema could be driven by outlier cases. Also, note that we are careful not to use the language that HH23 injected \gepbi\ into the \kepler\ data, since again the signal they injected was \textit{not} the signal proposed in the original paper but rather their own best-fit solution (${\neq}$\gepbi), which we have established to have missed the global optimum (i.e. \gepbi).

To test this claim, we drew random samples from the K22 joint posterior distribution for \gepbi\ and injected them into the \kepler\ data. Note that we do not assume Gaussian noise here; we use the actual \kepler\ light curves processed through our method marginalised detrending algorithm used by K22. From 22 experiments, we obtain a 50th percentile central range of $-1.2 \leq 2 \log B_{mp} \leq 9.2$, whereas the original detection was $(4.95\pm 0.59)$. Accordingly, our own analysis demonstrates that HH23's claim that the original Bayes factor was somehow surprising, given the proposed parameters, is also false.

\noindent\textit{\textbf{K1708: False-positive rate}}

HH23 state that ``the false positive rate of 1.6\% of our injection-retrieval tests suggests that an exomoon survey in a sufficiently large sample of transiting exoplanets with similar S/N characteristics yields a large probability of at least one false positive detection, which we think is what happened with Kepler-1708 b''. K22 devoted a section entitled ``Interpreting the FPP'' (in the context of the 70 objects surveyed), which addresses this point already. In summary, first, the FPP cannot be assumed to be 1\% for all 70 objects (we use K22's 1\% value since it's been established that HH23's light curve products are not equivalent to that used in the survey). Nevertheless, even if it \textit{were} 1\%, it trivially follows from the Binomial distribution that one would still be more likely to obtain 0 false-positives from a survey of 70 objects than 1. There's no legitimate way we can dismiss \gepbi\ as unlikely (or even refuted) when rigorously it's \textit{more} likely to not be a false-positive, than instead be spurious.

But truly, the probability that the signal is real is not merely one minus the FPP (which instead defines the true negative probability), but rather needs to fold in the prior probability of the moon being real. K22 show how it comes down a balancing act between the two. If your personal belief is that less than 1\% (the FPP) of \gepb\ analog planets have \gepbi\ analog moons, then you should not believe the moon candidate is real. If your prior is greater than 1\%, then you should believe it. This is really all the FPP can tell us. Without any exomoon catalog to guide us, the prior is at this point wholly subjective.

\noindent\textit{\textbf{K1708: No statistical evidence for red noise}}

HH23 write that ``The proposed exomoon transit signal is not distinct from other sources of variations in the light curve, which are probably of stellar or systematic origin.'' We point out this statement is immediately incompatible with their own injection-recovery tests. If the real \kepler\ time series is used for the injection, as was done in both K22 and HH23, then stellar and instrument variability is accounted for. Thus, the false-positive rate \textit{fully} encodes the chance that stellar variability conspired in such a way to create a moon-like dip. This emphasises the importance of avoiding chi-by-eye and the value rigorous statistical testing\cite{childs}. Planet+moon models are flexible, but they cannot explain \textit{anything}.

The argument that the exomoon signal is due to time correlated noise can be addressed another way besides from the false-positive rate, although we emphasise that the latter is far more relevant since it considers the entire end-to-end detection process. Regardless, we consider two standard statistical tests for time correlated noise, the Ljung-Box and Box-Pierce tests, as well as one for Gaussianity, the Kolmogorov–Smirnov test. For the two time-correlated noise tests, we explored lags up to 100 (i.e. 50\,hours) which is more than enough to cover the relevant timescale of the transit duration (19\,hours). In what follows, we report the the $p$-values from each test, where low values $(\lesssim 0.01)$ indicate that the data is inconsistent with white noise.

Taking the residuals from our best-fitting light curve reported in K22, for the first epoch, we find $p$-value of 0.88 for the Ljung-Box test, 0.96 for the Box-Pierce test and 0.27 for the Kolmogorov–Smirnov test. For the second transit, these become 0.52, 0.78 and 0.68 respectively.

In conclusion, neither the HH23 nor K22 reported false-positive rates, nor indeed do the standard Ljung-Box, Box-Pierce and Kolmogorov–Smirnov tests find any evidence for time-correlated noise being a likely explanation.

\noindent\textit{\textbf{K1708: Point-by-point rebuttal}}

\noindent We here address, point-by-point, the critiques raised in HH23 in the case of \gepbi.

\noindent\textit{
``(1) ...we also point out that at 1,508 d (BKJD), just about 1 d before transit 2, there was a substantial decrease in the apparent stellar brightness of ~800 ppm (see residuals in Fig. 3d,f) that is as deep as the suspected moon signal....''}
The probability that random fluctuations in the detrended light curve could conspire to mimic a signal that would be erroneously identified as a false-positive defines the false-positive rate, which is calculated in K22 and HH23 via injection-recovery tests of the real \kepler\ data. Both studies agree that the false-positive rate is ${\sim}1$\% and thus this possibility is fully accounted for in this calculation, but is clearly a remote possibility.

\noindent\textit{``(2) ...The proposed exomoon transit signal is not distinct from other sources of variations in the light curve, which are probably of stellar or systematic origin.''}
Again, the probability that stellar (or other) variability conspired in such a way to create a signal that would pass the tests used in K22 to identify exomoons is fully encoded within the false-positive rate, which equals $1.0_{-1.0}^{+0.7}$\%.

\noindent\textit{``(3) ...other variations in the phase-folded light curve that cannot possibly be related to a moon cast doubt on the exomoon hypothesis''}
Same as response \#2.

\noindent\textit{``(4) ...Due to geometrical considerations it is, in fact, unlikely a priori that a moon performs its own transit in a close apparent deflection to its planet.''}
The claim is that, even with just two transits, it's improbable that the moon would be temporally so close to the planet in both epochs. It should be noted that this statement is highly sensitive to the planet-moon semi-major axis. Close separations will of course lead to much shorter intervals between the planet and moon transit events. For \gepbi, the reported separation from K22 is indeed very close, approximately that of Europa around Jupiter. In any case, we anticipated this comment already in the original paper\cite{K22}. In our Supplementary Information Figure~12 of \cite{K22}, we evaluated the log-likelihood of obtaining the two moon transit times reported given the reported parameters. This was then compared to Monte Carlo simulations for random phases where one can see that the reported positions are almost precisely aligned with the median (i.e. typical) phasing expected. Ergo, the observed phase positions are not surprising. HH23 thus seem to be unaware that point \#4 has thus been specifically already shown to be false in the original paper.

\noindent\textit{``(5) Our orbital solutions for the proposed exomoon vary substantially depending on the detrending method... For a real and solid exomoon detection, we would expect that the solution is stable against various reasonable detrending methods.''}
The exomoon period is the most difficult parameter for any planet+moon fit to recover, evident even in injection recovery experiments of high S/N data\cite{luna}. However, the degree to which the exomoon period can be constrained should not be conflated with the weight of evidence in favour of the exomoon. To use an analogy, in direct imaging, a couple of images of a long-period planet will make it very difficult to constrain the orbital period, eccentricity and other orbital elements\cite{ferrer}. But the fact these parameters are difficult to constrain has no bearing on the reality of what a planet is in those images. Models often feature physical parameters that cannot be well-constrained but constraining those parameters, parameter inference, is a distinct question as to whether the physical object exists, model selection - and they should not be conflated.

\noindent\textbf{\large{Further Issues}}
\label{si:other}

\noindent\textit{\textbf{Ideal candidates for exomoon detectability}}

HH23 described ``idealized scenarios in which exomoons can be found'' by constructing a transit model for a Neptune-sized planet orbiting at Sun-like star in a $60$\,day orbit. These scenarios are fundamentally different than the actual conditions for \kepbi\ or \gepbi, as the exomoon candidate host planets orbit their host stars at much longer orbital periods. As described by HH23, they used such scenarios to identify conditions where more transits are possible. However, this is undermined by their selection of physical conditions that would not permit many of these exomoons to exist due to tidal interactions between both the host planet and star.

Exomoon orbits are limited to an inner (Roche limit) and outer boundary (Hill radius), where crossing these boundaries leads to either destruction or ejection, respectively, of the exomoon\cite{barnes}. HH23 described the Hill radius incorrectly for their proposed scenario and completely disregarded the Roche limit. Assuming a circular planetary orbit, the Hill radius must include an estimate of the exomoon mass because a Neptune-sized planet is used ($R_{\rm Hill} = a_P([M_P + M_M]/[3M_{\star}])^{1/3}$), which can increase the value of their $R_{\rm Hill}$ by ${\sim}2$\%. Additionally, more recent studies\cite{rosario,quarles2021} showed that the stability limit coefficients are ${\sim}20$-$30$\% smaller than those adopted in HH23 (which stem from reference \cite{domingos}) when considering a general initial condition of the exomoon’s orbit. By their metric for the Hill radius (or more recent estimates), the ``injection of an exomoon throughout the Hill radius'' leads to a large fraction of unstable scenarios (${\sim}82$\% of their trials if uniformly distributed by orbital period).

Tidal interactions make the inner boundary more important as the outer boundary cannot be reached\cite{sasaki,piro,quarles2020}. Since the host planet orbits much closer to the host star, the stellar tide despins the planet much more effectively, quickly limits the outward migration of the exomoon, and forces the exomoon to begin a much longer phase of in-fall (see Extended Data Figure~\ref{fig:tidal}). The maximum lifetime of their hypothetical system is ${\lesssim}5$\,Gyr and corresponds to a Type I: Case 1 solution in reference \cite{sasaki}. Note that our $y$-axis uses the more accurate version of the Hill radius given above. The exomoon’s lifetime can be extended to ${\sim}7.4$ \,Gyr if it begins at $0.17$\,$R_H$ instead of our assumed $0.1$\,$R_H$, however this only delays the inevitable in-fall. Starting beyond ${\sim}0.18$\,$R_H$ or reducing the host planet's initial spin rate results in a quick ejection of the exomoon (${<}1$\,Gyr) thereby increasing the percentage of unstable scenarios to ${\sim}95$\%.

The estimated ages of \kep\ and \gep\ are ${\sim}8$\,Gyr\cite{TK18} and ${\sim}3$\,Gyr\cite{K22}, respectively. The ages of Sun-like Kepler stars peak near ${\sim}3$\,Gyr, with a long tail toward older ages\cite{berger}, which implies that many of the ideal cases would have been tidally disrupted or nearing the Roche limit of their host planet. In conclusion, only ${\sim}5$\% of the exomoon injections result in a physical scenario where the exomoon could potentially be observed given the estimated stellar lifetimes and thus we disagree with HH23's claim that these represent ``idealized scenarios in which exomoons can be found''.

\noindent\textit{\textbf{Implementation errors with \cofiam}}

HH23 attempted a like-for-like comparison by using \cofiam\ to detrend the \kepler\ time series of \gepb. Two issues are present with this effort. First, the final light curve product used by K22 was not from \cofiam\ but rather from the method marginalisation of eight different approaches. Second, the authors state they use \wotan\cite{wotan} as the algorithmic implementation of \cofiam, but inspection of their code reveals numerous important differences to the version used in K22 (made available alongside that paper), which we list below.

\begin{enumerate}
\item \wotan's version doesn't account for the effect of the transit gap upon the Durbin-Watson (DW) statistic. The DW metric is only valid for regularly sampled data but after masking the transit (which is always necessary) the time series is highly irregular, featuring a large data gap where the transit once was. To account for this, \cofiam\ proper computes the DW separately on the pre- and post-transit data and then measures their combined deviance from 2 (uncorrelated noise) at the end. \wotan's version does not follow this recipe and thus will produce biased DW metrics.
\item \wotan's version does not weight the time series by the inverse square of the measurement uncertainties, and thus implicitly assumes homoskedastic data, unlike \cofiam\ proper. The light curve does exhibit small heteroskedasticity, but the effect is small enough that this is unlikely to have a major impact.
\item Jump times (times of temporal/flux discontinuity) are individually identified by the user in \cofiam\ proper, whereas \wotan\ relies upon a user-specified temporal gap to automatically flag them.
\item \cofiam\ proper runs through up to 30 orders for the sum of cosines function and selects the global optimum. \wotan's version monitors the DW and stops once the DW starts to increasingly deviate from 2. Thus it tends to pick a local optimum, rather than the global.
\end{enumerate}

\noindent\textit{\textbf{\luna\ versus \pandora\ speed comparison}}

HH23 claim that \pandora\cite{pandora}, the code used in HH23, is ``4 to 5 orders of magnitude'' faster than \luna\cite{luna}, as used in TK18 and K22. This would obviously be a truly enormous gulf in computational efficiency, if true. To test this claim, we took 1000 random samples from the \kepbi\ posterior samples reported in TK18 and computed the planet+moon model using both codes. We find that \pandora\ and \luna\ in fact have comparable run times, with the mean ratio of \luna\ to \pandora\ run-times being 1.33. The claim of ``4 to 5 orders of magnitude'' is certainly false and it's unclear to us how the authors arrived upon such an extreme value. 

\noindent\textit{\textbf{Misinterpretation of likelihood behaviour in posterior samples}}

We take the opportunity to address a criticism raised in H19 regarding the reality of \kepbi. In their Figure~3, the authors show two histograms of the log-likelihoods obtained from their posterior samples, one using the original TK18 data, and the other using their own reduction. In both cases, they suggest that the fact the mode of the distribution is not co-located with the maximum likelihood position is a problem, ``suggesting that the best-fit solution could in fact be an outlier''. We address this here because others have quoted this conclusion\cite{K19} and it represents a misinterpretation of the products from Bayesian inference.

In a one-dimensional problem, an MCMC walker will converge to a likelihood maximum (assuming uninformative priors) and oscillate around that position. Most MCMC samplers are renown as being poor tools to precisely determine the maximum likelihood solution, since by design they make finite steps that under/overshoot the likelihood maximum\cite{sharma}. Their objective is to sample the target distribution (the posterior) and not to iteratively hone in on the maximum likelihood solution. One can therefore see that the walker will spend most its time offset (but nearby) the likelihood maximum. As the dimensionality of the problem is increased, the walker obtains more degrees of freedom to step into, and this increased flexibility means it more easily deviates away from the maximum likelihood position, in a log-likelihood sense. This behaviour is well-known and quite expected and indeed has even been suggested as a useful tool\cite{trotta}.

It has been rigorously shown that effective dimensionality of a problem can even be measured from this deviance\cite{kunz}. In that work, the authors show that \textit{the effective number of parameters} equals

\begin{align}
\mathcal{C}_b &= \bar{\chi^2(\theta)} - \chi^2(\hat{\theta}),
\end{align}

where bar symbol denotes the mean over the posterior distribution, $\theta$ is a vector containing the model parameters and $\hat{\theta}$ is the maximum likelihood solution. An offset, as seen in H19's Figure~3, is thus precisely that which one would expect to see for a high dimensional problem such as this. Although we don't have their posteriors to precisely calculate $\mathcal{C}_b$, taking the TK18 posteriors which are publicly available along with that paper, we obtain $\mathcal{C}_b = 22.8$ for the quadratic visit-long trend model, comparable to the actual parameter volume dimensionality of 22.

In conclusion, H19's suggestion that the offset between mean and maximum likelihood is peculiar is inaccurate, and indeed this is in fact an established tool in Bayesian inference\cite{kunz}.


\clearpage
\begin{addendum}

\item[Data Availability]
The data that support the plots within this paper and other findings of this
study are available at \wwwcoolworlds; or from the corresponding author
upon reasonable request.

\item[Code Availability]
The \multi\ regression algorithm\cite{multi} is publicly available at
\href{https://github.com/farhanferoz/MultiNest}{github.com/farhanferoz/MultiNest}.
The \cofiam\ software package was released along with the original paper\cite{K22} at
\href{https://datadryad.org/stash/dataset/doi:10.5061/dryad.18931zcz9}{doi:10.5061/dryad.18931zcz9}.
The \luna\ forward model is publicly available at
\href{https://sourceforge.net/p/lunamod}{sourceforge.net/p/lunamod}.
The \satcand\ package\cite{quarles2020} is publicly available at
\href{https://github.com/Multiversario/satcand}{github.com/Multiversario/satcand}.

\end{addendum}



\newpage
\begin{landscape}
\begin{longtable}{lcccc}
\hline
Parameter & Median Intra-orbit RMS & Fit Residuals (lin) & Fit Residuals (quad) & Fit Residuals (exp) \\
\hline
Teachey \& Kipping 2018 (TK18) & 360.6\,ppm & 384.5\,ppm & 361.3\,ppm & 369.0\,ppm \\
Kreidberg et al. 2019 (K19) & 362.4\,ppm & 373.3\,ppm & 373.4\,ppm & 370.3\,ppm \\
Heller \& Hippke 2023 (HH23) [our fits] & 428.5\,ppm & 423.1\,ppm & 425.7\,ppm & 422.0\,ppm \\
\hline
\caption{Comparison of photometric precision obtained by three different reductions of the Hubble data for \kep. The second column ignores Hubble orbits 1, 7 and 18 and is a model independent measure of precision. Columns 3 to 5 jointly regress a visit-long trend model (linear, quadratic and exponential) and a planet+moon model to the various data sets and then measure the residual RMS.
}
\label{tab:K1625} 
\end{longtable}
\end{landscape}

\newpage
\begin{longtable}{lcccc}
\hline
Parameter & Planet-fit RMS [ppm] & Moon-fit RMS [ppm] \\
\hline
Heller \& Hippke\cite{HH23} reported & 529.9 & 528.2 \\
Heller \& Hippke\cite{HH23} extraction [alt. fit$^{\dagger}$] & 529.3 & 510.4 \\
Heller \& Hippke\cite{HH23} extraction & 527.8 & 507.9 \\
\wotan\cite{wotan} + bifilter & 516.1 & 506.3 \\
Original \gepbi\ article\cite{K22} & 495.9 & 491.1 \\
\hline
\caption{Comparison of photometric precision obtained by different detrending methods of the \kepler\ data for \gep. The alternative fit, marked by $^{\dagger}$, corresponds to a fit where the out-of-transit baseline is forced to unity (assumes perfecting detrending). In general, the fits of this work do not make this assumption but it is adopted here to provide a fair like-to-like comparison with Heller \& Hippke\cite{HH23}.
}
\label{tab:K1708} 
\end{longtable}

\newpage
\begin{efigure}
  \centering
  \includegraphics[angle=0, width=16.0cm]{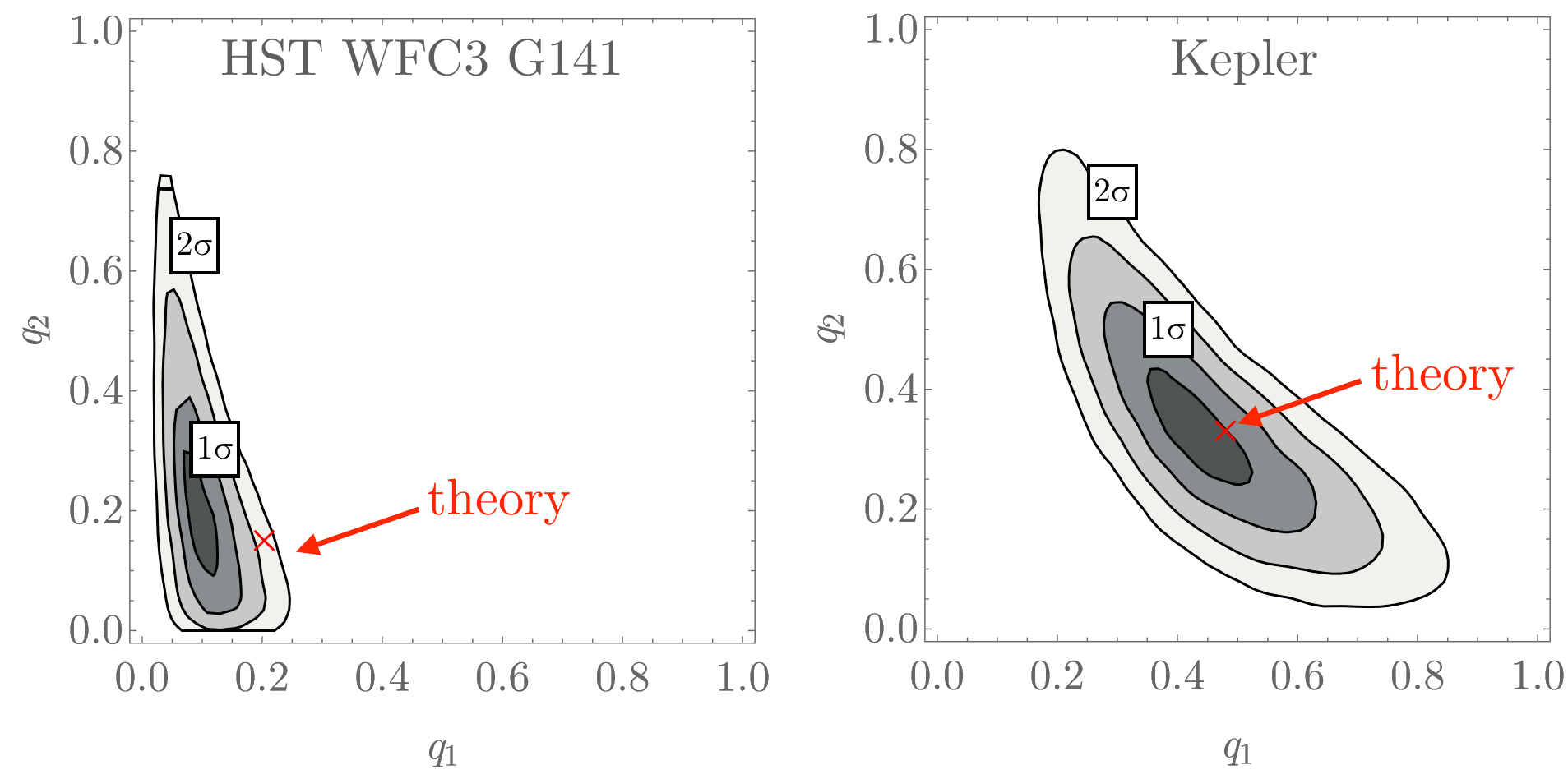}
  \caption{\label{fig:LDCs}
  \textbf{Joint posterior distributions for the limb darkening coefficients of \kep\ taken from \cite{TK18}.} We mark the theoretical expectation with a red cross, computed using the Kurucz stellar atmosphere database and described in the SI. The original limb darkening treatment appears consistent with astrophysical models.
  }
\end{efigure}

\newpage
\begin{efigure}
  \centering
  \includegraphics[angle=0, width=16.0cm]{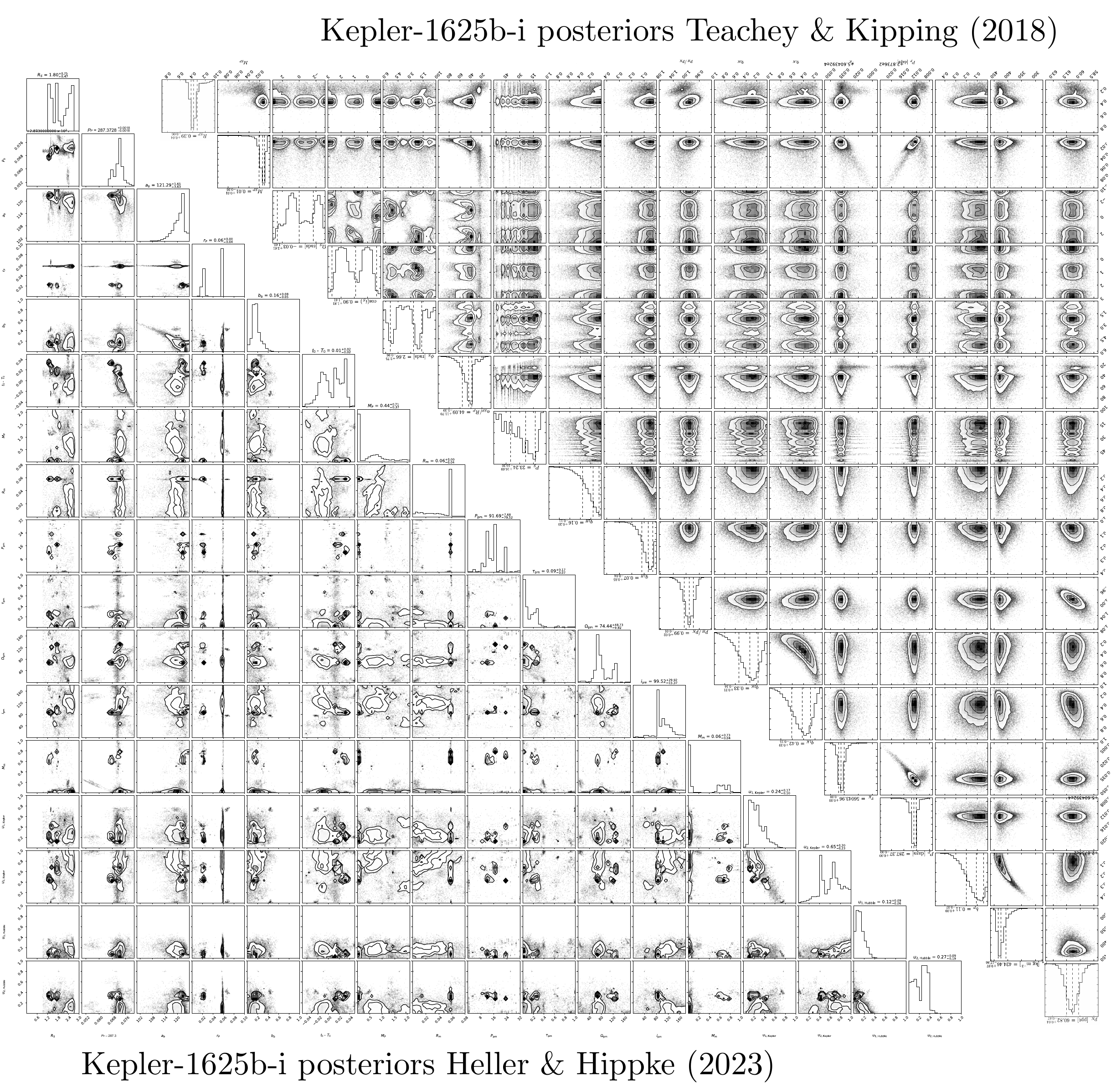}
  \caption{\label{fig:corners}
  \textbf{Comparison of the posterior corner plots for \kepbi\ from \cite{TK18} and \cite{HH23}.}
  Comparison shows the two corner plots lifted from published articles. The only change we make is
  that the corner plot from \cite{TK18} has been mirrored in both directions to let both plots
  fit on the page.
  }
\end{efigure}

\newpage
\begin{efigure}
  \centering
  \includegraphics[angle=0, width=16.0cm]{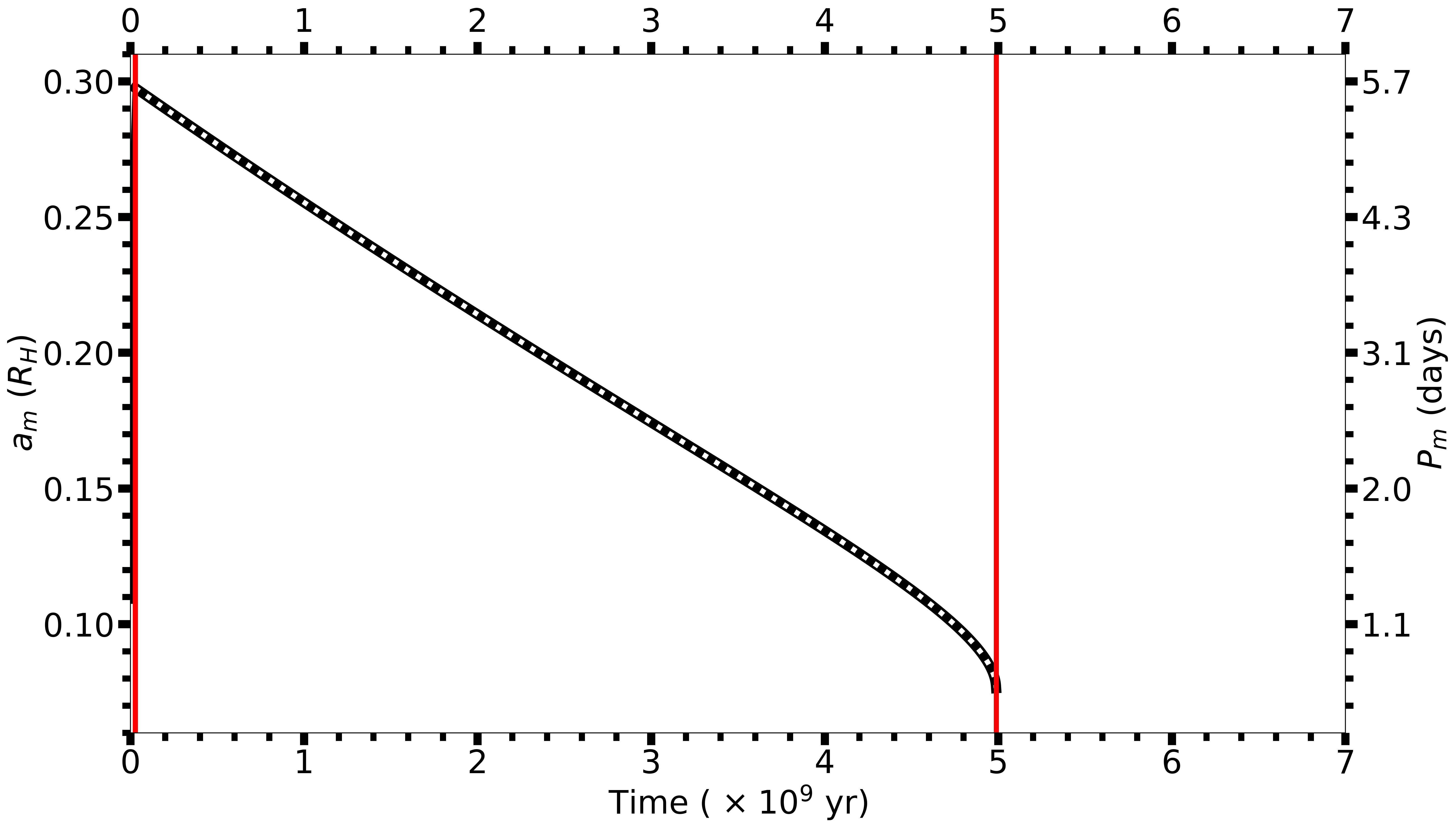}
  \caption{\label{fig:tidal}
  \textbf{Tidal evolution of an Earth-sized exomoon orbiting a Neptune-size planet around a Sun-like star using \satcand\cite{quarles2020}.}  The exomoon begins at $5$\,$R_P$ (or $0.11$\,$R_H$), while the initial planetary rotation period is $6$\,hr.  All other parameters follow those given in reference \cite{sasaki}; see their Figure 14.
  }
\end{efigure}

\end{document}